\documentclass[%
aip,
amsmath,amssymb,
reprint,%
]{revtex4-1}
\usepackage{graphicx}% Include figure files
\usepackage{dcolumn}% Align table columns on decimal point
\usepackage{bm}% bold math
\usepackage[utf8]{inputenc}
\usepackage[T1]{fontenc}
\usepackage{mathptmx}
\usepackage{etoolbox}
\usepackage{xcolor}

\newcommand{\GW}{$G_0W_0$}

\makeatletter
\def\@email#1#2{%
	\endgroup
	\patchcmd{\titleblock@produce}
	{\frontmatter@RRAPformat}
	{\frontmatter@RRAPformat{\produce@RRAP{*#1\href{mailto:#2}{#2}}}\frontmatter@RRAPformat}
	{}{}
}%
\makeatother
\begin{document}
	
	\preprint{AIP/123-QED}
	
	\title[InAlN nanorods]{Electronic and optical properties of core-shell InAlN nanorods: a comparative study via LDA, LDA-1/2, mBJ and \GW{} methods }
	% Force line breaks with \\
	\author{Ronaldo Rodrigues Pela}
	\affiliation{Supercomputing Department, Zuse Institute Berlin (ZIB), Takustraße 7,	14195 Berlin, Germany}
	\email{ronaldo.rodrigues@zib.de}
	\author{Ching-Lien Hsiao}
	\author{Lars Hultman}
	\author{Jens Birch}
	\author{Gueorgui Kostov Gueorguiev}
	%	\homepage{http://www.Second.institution.edu/~Charlie.Author.}
	%	\email{Second.Author@institution.edu.}
	\affiliation{ 
		Thin film Physics Division, Department of Physics, Chemistry and Biology (IFM), Linköping University, SE 581 83 Linköping,
		Sweden%\\This line break forced with \textbackslash\textbackslash
	}%
	
	\date{\today}% It is always \today, today,
	
	\begin{abstract}
		Currently, self-induced InAlN core-shell nanorods enjoy an advanced stage of accumulation of experimental data from their growth and characterization as well as a comprehensive understanding of their formation mechanism by the \textit{ab initio} modeling based on Synthetic Growth Concept. 
		However, their electronic and optical properties, on which most
		of their foreseen applications are expected to depend, have not been investigated comprehensively. 
		\GW{} is currently regarded as a gold-standard methodology with quasi-particle corrections to calculate electronic properties of materials in general. 
		It is also the starting point for higher-order methods that study excitonic effects, such as those based on the Bethe-Salpeter equation.
		One major drawback of \GW{}, however, is its computational cost, much higher than density-functional theory (DFT).
		Therefore, in many applications, it is highly desirable to answer the question of how well approaches based on DFT, such as \textit{e. g.} LDA, LDA-1/2, and mBJ, can approximately reproduce \GW{} results with respect to the electronic and optical properties. 
		Thus, the purpose of the present paper is to investigate how the DFT-based methodologies LDA, LDA-1/2, and mBJ can be used as tools to approximate \GW{} in studies of the electronic and optical properties of scaled down models of core-shell InAlN nanorods.
		For these systems, we observed that band gaps, density of states, dielectric functions, refractive indexes, absorption and reflectance coefficients are reasonably well described by LDA-1/2 and mBJ when compared to \GW{}, however, at a much more favorable computational cost.
		
	\end{abstract}
	
	\maketitle

\section{Introduction}

Wurtzite InAlN semiconductor alloys have a direct band gap that span a wide spectrum range from 0.65 eV (InN) to 6.25 eV (AlN). \cite{Taniyasu_2006,Wu_2002,Hsiao_2007}
Therefore, many optoelectronic devices can possibly be fabricated from InAlN alloys, which are applicable in a wide wavelength range covering deep-ultraviolet (DUV) to near infrared (NIR), such as light-emitting diodes, laser diodes, solar cells, and photodetectors.\cite{Baten_2021,Alias_2018,Huang_2019,Li_2017b,Filho_2023}
However, InAlN thin film often contains large number of structural defects and compositional inhomogeneity owing to a wide-range composition immiscibility of the In$_x$Al$_{1-x}$N ($0.1 < x < 0.9$), low dissociation temperature of InN ($\sim 550$ °C), and mismatches in lattice and coefficient of thermal expansion to common substrates.\cite{Nanishi_2003,Ferhat_2002,Palisaitis_2017} 
Alternatively, InAlN grown in the form of low-dimensional nanostructures can provide an opportunity to overcome the effects of lattice mismatch like threading dislocations formation and substrate-film strain. 
\par In the context of the InAlN low-dimensional nanostructures, self-induced core-shell InAlN nanorods (NRs) have been successfully synthesized by reactive magnetron sputter epitaxy (MSE) while their formation mechanism was elucidated by modeling the relevant precursor prevalence and their corresponding energetics using the DFT-based synthetic growth concept (SGC).\cite{Filho_2023}
SGC is an extensive approach designed for modeling of nanostructures with complex morphology and accounting for the role of the precursors in their formation when employing a wide spectrum of vapor-phase deposition techniques.\cite{Goyenola_2012,Gueorguiev_2011,Gueorguiev_2006,Goyenola_2011}
\par Very high-crystal-quality nanomaterials can be grown on various substrates, including metals, metal nitrides, oxides and Si,\cite{Kamimura_2007,Prabaswara_2020,Serban_2017} which opens the possibility of integration with mature device-fabrication technology for integrated circuit industry. 
Furthermore, the form of nanostructure enables to fabricate nanodevices with high performance benefited from the reduced geometry. 
For instance, InAlN nanospirals with tailored chirality have been demonstrated to reflect circularly polarized light with corresponding handedness, through tuning internally compositional distribution and external geometry, which is very promising for fabricating high-performance optical elements.\cite{Hsiao_2015,Kuo_2018} High-sensitivity photodetectors based on InAlN nanophotonic structure is applicable from deep DUV to NIR region.\cite{Yao_2017,Alias_2018,Li_2017b}
With controlling composition of InAlN with In-content $\sim 0.17$, strain-less multilayer InAlN/GaN-distributed Bragg reﬂectors (DBRs) with high a peak reflectivity can be grown directly onto nanodevice’s structures for fabricating vertical-cavity surface-emitting lasers (VCSELs).\cite{Ristic_2005,Takeuchi_2018}

\par To aid the development of nanodevices based on core-shell InAlN NRs, it is crucial to have a theoretical tool to test different design scenarios and to help the
interpretation of the electronic properties of as-synthesized core-shell InAlN NRs. 
Reliable simulation of their optical properties provides a strategic tool for tuning the core-shell InAlN NRs to potential electronic and optoelectronic applications.
In this sense, it is desirable a methodology that accurately describes the excitations of such nanostructures across a wide energy range, especially around the bandgap-energy region. The solution to this problem is given by the \GW{} approximation within many-body perturbation theory, which is considered the state of the art in \emph{ab initio} calculations of single-particle excitations.\cite{Aryasetiawan_1998,Onida_2002,Bechstedt, Golze_2019,Reining_2018} It can provide accurate quasi-particle corrections to (generalized) Kohn-Sham eigenvalues, yielding electronic structures in excellent agreement with experiments and with higher-order methods.\cite{Nabok_2016,Bechstedt,Govoni_2015,vanSetten_2015,Krause_2015} However, a major drawback of \GW{} is its high computational cost, which can complicate its application to complex systems with hundreds or thousands of atoms.\cite{Pela_2015}

\par For this purpose, it is interesting to find approaches based on DFT that can reproduce \GW{} results, with reasonable accuracy, but much less computationally involved. Among the various possibilities, in this paper, we explore two: LDA-1/2 and the modified Becke-Johnson (mBJ) functional. 

\par The LDA-1/2 approach has proven to be an efficient alternative for obtaining approximate quasi-particle corrections at low computational cost. \cite{Ferreira_2008,Ferreira_2011,Pela_2012,Pela_2016,Pela_2018,Matusalem_2013,Pela_2011,Pela_2015}
In particular, electronic properties of systems based on III-V semiconductors are well described by LDA-1/2.\cite{Pela_2011,Pela_2015,SilvaFilho_2013,Santos_2012} For this class of materials, LDA-1/2 also provides accurate one-particle energies and wavefunctions to solve the Bethe-Salpeter equation and obtain optical properties.\cite{Matusalem_2020} 
Regarding nanowires, LDA-1/2 calculations for Si, GaN, and GaP have been shown to describe the band gap with an accuracy comparable to \GW{}\cite{Ribeiro_2015,Huang_2015} and in good agreement with experiments.\cite{Greil_2016} 
These facts make LDA-1/2 an attractive \emph{ab initio} framework to study core-shell InAlN NRs. To what extent this is possible, however, has not yet been addressed.

\par Another promising choice is the mBJ potential,\cite{Tran_2019,Tran_2009} a semilocal meta-GGA functional shown to be quite accurate for band gap calculations. It is competitive with \GW{} and hybrid functionals in terms of accuracy, at much lower computational cost.\cite{Koller_2011,Kim_2010,Tran_2019,Jiang_2013,Lee_2016} 
Interestingly, band gaps of III-V semiconductors calculated with mBJ show good agreement with experiments.\cite{Wang_2013,Kim_2010} 
Apart from band gaps, optical properties of several materials have been obtained with mBJ,\cite{Li_2017,Ibarra_2017,Roedl_2019,Singh_2010,Ondracka_2017,Nakano_2018}
including III-V semiconductors,\cite{Rehman_2016,Nakano_2018} and mBJ at least improves over PBE when compared with experiment.\cite{Nakano_2018}
Studies employing mBJ for nanowires have been conducted as well,\cite{Radzwan_2019,Xiang_2017,Park_2022,Validvzic_2014}
some of which have reported nice agreement with measurements.\cite{Park_2022,Validvzic_2014}
It it, thus, important to verify how mBJ performs for studying core-shell InAlN NRs.

\par In this work, for the case of core-shell InAlN NRs, we conduct \emph{ab initio} calculations to analyze how LDA-1/2 and mBJ improve over LDA and how they can approximate \GW{} for the following electronic and optical properties: density of states (DOS), band gaps, dielectric function, refraction index, extinction and absorption coefficients, and the reflectivity.
Nanostructures of similar structural and chemical complexity and their electronic and optical properties including in relation to electronic applications have been successfully studied previously by using both different flavors of GGA to DFT levels of theory \cite{Gueorguiev_2008}, and the \GW{} method. \cite{Oliveira_2021}
Here, to keep the computational cost moderate in the \GW{} calculations, we select as prototypes NRs with dia\-me\-ter of 14 \AA{} and with In compositions of $0$, $12.5$, and $25$\% within their core. 

\par The paper is divided as follows: in section \ref{sec:theory}, we introduce the theoretical aspects of this work; section \ref{sec:methods} describes the computational methods employed; in section \ref{sec:results}, we present and discuss our results; and lastly, in section \ref{sec:conclusions}, we summarize the paper.

\section{Theoretical Framework}\label{sec:theory}
\subsection{The LDA-1/2 method}
LDA-1/2\cite{Ferreira_2008,Ferreira_2011,Mao_2022} is inspired on Slater's half-occupation scheme, which relates the ionization potential $I$ of a KS eigenstate labeled with $i$ at its eigenvalue $E_i$:
\begin{equation}
I = -E_i(f_i=1/2),
\end{equation}
where $f_i$ is the occupation of the KS state $i$.

In LDA-1/2, instead of dealing with half-occupations, KS equations are modified as:
\begin{equation}
\left[
-\frac{1}{2}\nabla^2 + 
V_{H}(\mathbf{r})+V_{XC}(\mathbf{r})+V_S(\mathbf{r})
\right]
\phi_{i\mathbf{k}}(\mathbf{r}) 
= E_{i\mathbf{k}} \phi_{i\mathbf{k}}(\mathbf{r}).
\end{equation}
Here, we consider electrons in a solid with wavevector given by $\mathbf{k}$. $ \phi_{i\mathbf{k}}$ is the corresponding KS wavefunction. The KS potential, $V_{KS}(\mathbf{r}) = V_{H}(\mathbf{r})+V_{XC}(\mathbf{r})$, written as the sum of the Hartree, $V_{H}(\mathbf{r})$, and the exchange-correlation (XC), $V_{XC}(\mathbf{r})$, potentials has been adjusted to include $V_S(\mathbf{r})$, the so-called self-energy potential.\cite{Ferreira_2008} The XC potential employed here is LDA.\cite{Kohn_1965} For each atom in the solid, $V_S(\mathbf{r})$ is obtained from two calculations with the isolated atom as
\begin{equation}\label{eq:VS}
V_S(\mathbf{r}) = \Theta(\mathbf{r})[V_{KS,atom}(\mathbf{r})_{f_i=1/2}-V_{KS,atom}(\mathbf{r})_{f_i=1}],
\end{equation}
in which, we add an extra label $f_i$ to $V_{KS}$ to denote the occupation. $\Theta(\mathbf{r})$ is a trimming function to avoid the divergence due to the tail $1/(2r)$ coming from the difference of the two KS potentials in (\ref{eq:VS}). Historically, $\Theta(\mathbf{r})$ has been chosen as
\begin{equation}
\Theta(\mathbf{r})=\left\{
\begin{array}{cr}
\left[1-\left(\frac{r}{R_{CUT}}\right)^8\right]^3, & r\le R_{CUT},\\
0,& r>R_{CUT},
\end{array}
\right.
\end{equation}
where $R_{CUT}$ is the cutoff radius, which is determined variationally\cite{Ferreira_2008} and has proven to be transferable among different systems.\cite{Ferreira_2011}

\subsection{mBJ}
The mBJ potential keeps the correlation potential the same as in LDA and replaces the exchange potential with:\cite{Tran_2009,Tran_2019}
\begin{equation}\label{eq_mbj}
v_{x,\sigma}^{mBJ}(\mathrm{r}) = cv_{x,\sigma}^{BR}(\mathrm{r}) + (3c-2)\frac{1}{\pi}\sqrt{\frac{5}{6}}\sqrt{\frac{t_\sigma(\mathbf{r})}{\rho_\sigma(\mathbf{r})}},
\end{equation}
where $\rho_\sigma(\mathbf{r})$ is the density of electrons with spin $\sigma$, $t_\sigma(\mathbf{r})$ is the corresponding kinetic-energy density, and 
$v_{x,\sigma}^{BR}(\mathrm{r})$ is the Becke-Roussel potential\cite{Becke_1989}. The factor $c$ in (\ref{eq_mbj}) is evaluated as\cite{Tran_2009}
\begin{equation}
	c = \alpha + \beta\sqrt{\frac{1}{2\Omega}\int_\Omega \mathrm{d}\mathbf{r} \left[ \frac{|\nabla \rho_\uparrow(\mathbf{r})|}{\rho_\uparrow(\mathbf{r})} + \frac{|\nabla \rho_\downarrow(\mathbf{r})|}{\rho_\downarrow(\mathbf{r})} \right] },
\end{equation}
where $\alpha=-0.012$ and $\beta=1.023$~bohr$^{1/2}$, and $\Omega$ is the volume of a unit cell.

\subsection{\GW{} approach}
Taking KS eigenvalues and wavefunctions as reference, quasiparticle-corrected eigenvalues $E_{i\mathbf{k}}^{QP}$ can be calculated in the \GW{} approximation as:\cite{Rangel_2020,Reining_2018,Golze_2019,Onida_2002}
\begin{equation}
E_{i\mathbf{k}}^{QP} = E_{i\mathbf{k}} + Z_{i\mathbf{k}}
\{
\mathrm{Re}[\Sigma_{i\mathbf{k}}(E_{i\mathbf{k}})]-V_{XC,i\mathbf{k}}\},
\end{equation}
where $Z_{i\mathbf{k}}$ is the quasiparticle renormalization factor, and $\Sigma_{i\mathbf{k}}(\omega)$ and $V_{XC,i\mathbf{k}}$ are matrix elements of the self-energy ($\Sigma(\mathbf{r},\mathbf{r}',\omega)$) and the exchange-correlation potential:
\begin{equation}
	\Sigma_{i\mathbf{k}}(\omega)= \int \mathrm{d}\mathbf{r}\mathrm{d}\mathbf{r}'
	\phi^*_{i\mathbf{k}}(\mathbf{r})
	\Sigma(\mathbf{r},\mathbf{r}',\omega) \phi_{i\mathbf{k}}(\mathbf{r}'),
\end{equation}
\begin{equation}
	V_{XC,i\mathbf{k}}= \int \mathrm{d}\mathbf{r}
	V_{XC}(\mathbf{r}) |\phi_{i\mathbf{k}}(\mathbf{r})|^2.
\end{equation}
Within the \GW{} approximation, the self-energy $\Sigma(\mathbf{r},\mathbf{r}',\omega)$ is given, in the time domain, as a product of the imaginary number, the single particle Green's function, $G_0(\mathbf{r},\mathbf{r}',t)$, and
the screened Coulomb interaction, $W_0(\mathbf{r},\mathbf{r}',t)$, evaluated in the random-phase approximation.\cite{Reining_2018,Freysoldt_2007}

\subsection{Optical properties}
Neglecting excitonic effects and considering an electric field applied along the $\hat{\mathbf{e}}_{\alpha}$ direction, the tensorial component $\alpha \alpha$ of the dielectric function $\varepsilon$, at a given frequency $\omega$, has an imaginary part given by\cite{Bassani}:
\begin{equation}\label{eq:im_epsilon}
	\mathrm{Im}[ \varepsilon_{\alpha\alpha}(\omega) ] = \frac{8 \pi^2}{\Omega N_\mathbf{k}} \sum_{cv\mathbf{k}} 
	\frac{|\langle \phi_{c\mathbf{k}} | -\mathrm{i}\hat{\mathbf{e}}_{\alpha} \cdot \mathbf{\nabla} | \phi_{v\mathbf{k}} \rangle|^2}{\omega^2} 
	\delta(\omega - \omega_{cv\mathbf{k}}),
\end{equation}
where $N_\mathbf{k}$ is the number of \textbf{k}-points, $c$ and $v$ are labels for the conduction and valence states, respectively, and $\phi_{c\mathbf{k}}$ and $\phi_{v\mathbf{k}}$ are the corresponding KS wavefunctions. The transition energies, $\omega_{cv\mathbf{k}}$, are expressed in terms of the KS eigenvalues as:
\begin{equation}
	\omega_{cv\mathbf{k}} = E_{c\mathbf{k}} - E_{v\mathbf{k}}.
\end{equation}

If the imaginary part is known, the real part can be obtained using the Kramers-Kronig relations\cite{Cardona}:
\begin{equation}\label{eq_re_epsilon}
	\mathrm{Re}[ \varepsilon_{\alpha\alpha}(\omega) ]
	=
	1 + \frac{2}{\pi}\int_0^\infty \mathrm{d}\omega'
	\frac{\omega'\mathrm{Im}[ \varepsilon_{\alpha\alpha}(\omega')]}{\omega'^2-\omega^2}.
\end{equation}
With $\epsilon_{\alpha\alpha}(\omega)$, it is possible to obtain other optical properties, such as the refraction index $\tilde{n}$, the extinction coefficient $\kappa$, the optical absorption $\mathcal{A}$ and the reflectivity $\mathcal{R}$\cite{Dresselhaus}:
\begin{equation}\label{eq_refraction_index}
	\tilde{n}(\omega) = \sqrt{\frac{|\varepsilon(\omega)|+\mathrm{Re}[ \varepsilon(\omega)]}{2}},
\quad
	\kappa(\omega) = \sqrt{\frac{|\varepsilon(\omega)|-\mathrm{Re}[ \varepsilon(\omega)]}{2}},
\end{equation}
\begin{equation}\label{eq-absorption-coeff}
\mathcal{A}(\omega) = \frac{2\omega \kappa}{v_{light}},
\quad
\mathcal{R}(\omega) = \frac{(1-\tilde{n})^2+\kappa^2}{(1+\tilde{n})^2+\kappa^2},
\end{equation}
where $v_{light}$ is the light speed in vacuum. For simplicity, we dropped down the double indexes $\alpha\alpha$ in Eqs. (\ref{eq_refraction_index}) and (\ref{eq-absorption-coeff}).

\section{Computational methods}\label{sec:methods}
In all DFT calculations, we employ the Quantum Espresso code\cite{Giannozzi_2009,Giannozzi_2017,Giannozzi_2020} with optimized norm-conserving Vanderbilt pseudopotentials\cite{Hamann_2012} and a planewave cutoff of 100 Ry. For the \GW{} calculations, we make use of BerkeleyGW\cite{Deslippe_2012,Hybertsen_1986}, taking LDA as the starting-point. We take advantage of the static remainder approach\cite{Deslippe_2013} to speed up convergence with respect to the unoccupied states. To reduce the computational cost, we use the plasmon-pole approximation.\cite{Zhang_1989,Hybertsen_1986}

We start the study with bulk AlN and InN in the wurtzite phase. We employ the experimental lattice parameters\cite{Vurgaftman_2003}, relaxing the ions positions with LDA. Then, the same relaxed geometry is used for all other methods. We use a k-grid of $16\times 16\times 10$ for LDA, LDA-1/2 and mBJ. For the \GW{} calculations, we consider an extrapolation scheme, as described in Appendix \ref{appendix:convergence-bulk}: we use k-grids of $4\times 4\times 3$ and $8\times 8\times 6$, and vary the cutoff for the dielectric function from 30 to 60 Ry in steps of 10 Ry, and the number of KS states from 100 to 450 in steps of 50. 

\begin{figure}[htb]
	\includegraphics[scale=0.2]{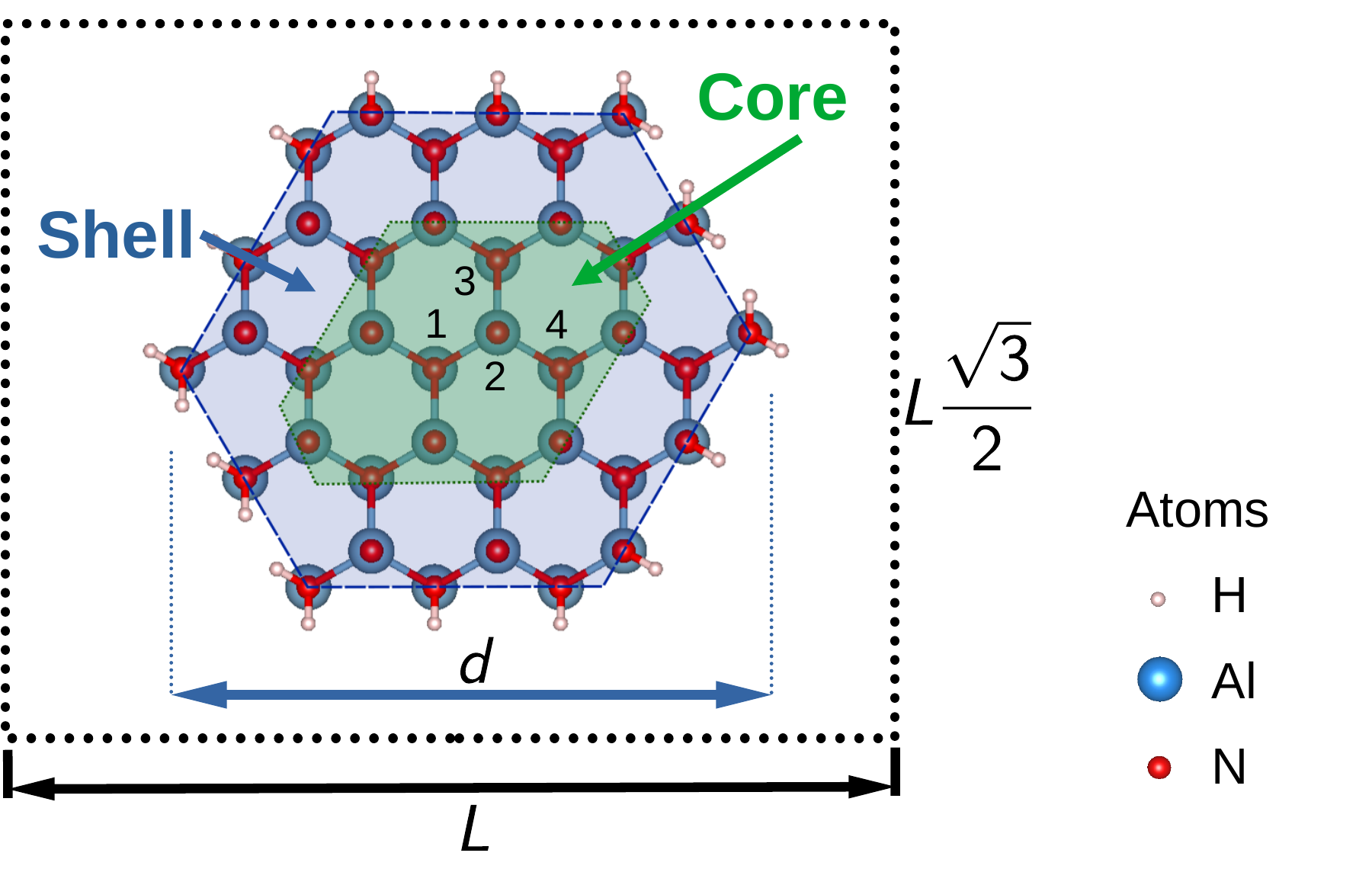}	
	\caption{Cell used to accommodate the passivated core-shell InAlN NRs (In$_n$Al$_{38-n}$N$_{38}$H$_{40}$). The shown diameter, $d$, is 14 \AA. We employ $L=44$ Bohr (23.3 \AA). We depict here the case of an AlN NR ($n=0$). In the case $n=2$, Al atoms at sites labeled with 1 and 2 are replaced by In atoms. In the case $n=4$, then all 4 labeled sites are replaced by In atoms. }\label{fig:Shell-Core}
\end{figure}

Then, we proceed to the core-shell InAlN NRs. We take as diameter $d=14$~\AA, as illustrated in Fig. \ref{fig:Shell-Core}. 
Even though much larger cells may be required to study realistic NRs,\cite{Filho_2023} our goal here is to evaluate the accuracy of LDA-1/2 and mBJ in approximating \GW{} for these systems. Keeping the computational cost of \GW{} in mind, we selected these NRs with a relatively small diameter as prototypes for our benchmark. We rationalize that with these NRs, one is still able to draw meaningful conclusions. Then further studies can then profit from our analysis and employ LDA-1/2 or mBJ to investigate NRs with more realistic sizes.

To avoid dangling bonds which lead to spurious states at the Fermi energy, we use H passivation, and, so, the chemical formula of the NR becomes In$_n$Al$_{38-n}$N$_{38}$H$_{40}$. We study three different In concentrations: $n=0$, 2, and 4. In all cases, we consider an unrelaxed geometry with bond lengths determined from the AlN experimental lattice parameters.\cite{Vurgaftman_2003} In Fig. \ref{fig:Shell-Core}, we also show a possible split between core and shell regions, leading to the compositions In$_n$Al$_{16-n}$N$_{16}$ for the core, and Al$_{22}$N$_{22}$H$_{40}$ for the shell. According to this choice, the cases $n=2$ and 4 correspond to In compositions of 12.5 and 25 \% in the core.
To isolate neighboring NRs, we employ a supercell with dimensions 44 Bohr $\times$ 31.1~Bohr ($23.3$~\AA $\times $ 20.2~\AA). 

For \GW{}, we employ a k-grid of $1\times 1 \times 16$ to obtain the reference density with LDA, and then $1\times 1 \times 6$ to generate the reference KS wavefunctions and eigenvalues. To enable a fair comparison, the same procedure is adopted for LDA, LDA-1/2 and mBJ. For the DOS and the optical properties, we take for all methods 300, 320, and 400 KS states into account, which is sufficient to cover transitions in the energy range $0-20$ eV.
In the \GW{} calculations, we include 900, 920, 940 bands in the summation used to build the dielectric function, with a cutoff of 20 Ry. To speed up the \GW{} convergence with respect to the vacuum size, we employ a Coulomb truncation for nanowires.\cite{Ismail-Beigi_2006}

\section{Results}\label{sec:results}
\subsection{Binaries: AlN and InN}
The purely binary compositions, AlN and InN, can be seen as benchmarks in relation to the ternary InAlN compounds and their properties are of relevance to the present first-principles comparative study of the InAlN core-shell NRs study.
\subsubsection{DOS and band gaps}
In Table \ref{tab:gaps-binaries}, the calculated band gaps of AlN and InN are compared with the experimental ones. 
As usual, LDA band gaps are underestimated for both AlN and InN. With LDA-1/2, although the band gap of InN is overestimated by 0.60 eV, the band gap of AlN agrees with the experimental with an error of 0.04 eV. With mBJ, in contrast, the band gap of InN deviates from experiment by 0.22 eV, while this error is 0.55 eV for AlN. Band gaps obtained with \GW{} agree with experiment with an error of 0.04 eV for AlN, and of 0.50 eV for InN. Overall, there is a similar degree of agreement with experiment for LDA-1/2, mBJ and \GW.
\begin{table}[htb]
	\caption{Calculated band gaps, compared with experimental gaps taken from Ref. \onlinecite{Vurgaftman_2003}.}\label{tab:gaps-binaries}
	\begin{tabular}{cccc}
		\hline
		& AlN & InN\\ \hline
		\GW{} & 6.29 & 0.28\\
		LDA & 4.24 & $-0.23$ \\
		LDA-1/2 & 6.21 & 1.38 \\
		mBJ &  5.70 & 0.56 \\
		exp. & 6.25 & 0.78\\
		\hline
	\end{tabular}
\end{table}

\begin{figure}[htb]
	\includegraphics[scale=1]{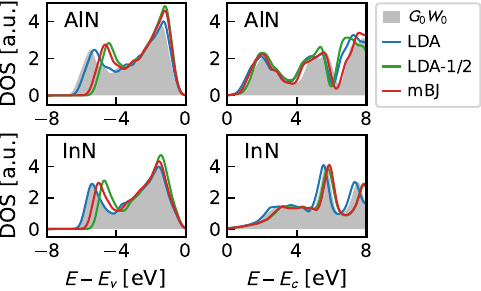}	
	\caption{DOS of AlN and InN. States belonging to valence and conduction bands are plotted on the left and on the right, respectively. In each case, band edges are placed at zero.}\label{fig:DOS-binaries}
\end{figure}

Figure \ref{fig:DOS-binaries} depicts the DOS of AlN and InN. 
For an easier comparison of the approaches, we plot the DOS of valence and conduction bands separately, on the left and on the right, respectively, placing in each case the band edges at zero. It is apparent then that LDA and \GW{} have the best agreement, confirming for AlN and InN the common belief that \GW{} approximately shifts states rigidly. DOS obtained with LDA-1/2 and mBJ agree well with each other and are also very close to \GW.

\subsubsection{Dielectric function}
\begin{figure}[htb]
	\includegraphics[scale=1]{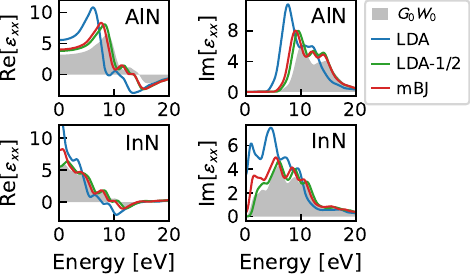}	
	\caption{$xx$ component of the dielectric function: left, the real part, and right, the imaginary part.}\label{fig:epsilon-binaries}
\end{figure}

We present the $xx$ component of dielectric function in Fig. \ref{fig:epsilon-binaries}. 
For AlN, the dielectric functions computed with LDA, LDA-1/2 and mBJ are red-shifted when compared to \GW, with LDA showing the largest deviation, and LDA-1/2 presenting a slightly better agreement than mBJ. For InN, the negative gap obtained with LDA causes a qualitative wrong behavior of $\varepsilon$ for small frequencies. LDA-1/2 and mBJ show similar results, with LDA-1/2 closer to \GW.

\subsection{Core-shell InAlN NRs}
\subsubsection{DOS}
\begin{figure}[htb]
	\includegraphics[scale=1]{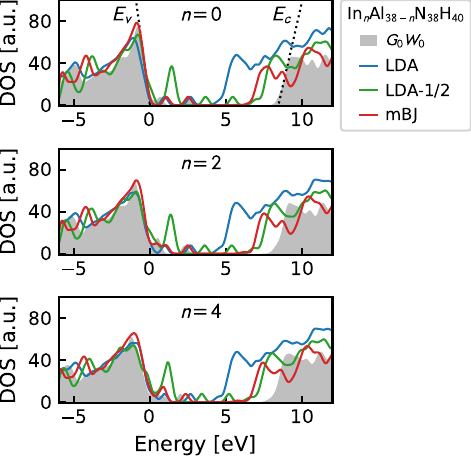}	
	\caption{DOS of passivated core-shell InAlN NRs for different In compositions. The dashed line on the top panel shows how band edges have been evaluated to obtain band gaps shown in Table \ref{tab:gaps-nanorods}.}\label{fig:DOS-nanorods}
\end{figure}

Fig. \ref{fig:DOS-nanorods} displays the DOS of core-shell InAlN NRs passivated with hydrogen. 
In each case, the zero energy has been defined as follows:
\begin{enumerate}
	\item projecting the DOS onto core atoms;
	\item identifying the valence state with the highest energy $E_v$;
	\item taking $E_v$ as reference and referring all other energies with respect to it.
\end{enumerate}
The identification of $E_v$ is illustrated for \GW{} DOS in the top panel of Fig. \ref{fig:DOS-nanorods} with the dashed line on the left. Similarly, we can define $E_c$ by projecting the DOS onto core atoms, and taking it as the energy of the conduction band edge. This is shown for \GW{} as the dashed line on the right in Fig. \ref{fig:DOS-nanorods} (subplot on the top).

\par The isolated peaks observed for energies between 0-5 eV come from states belonging to shell atoms.
The agreement between LDA, LDA-1/2, and mBJ with \GW{} for valence states with $E<0$ is evident for the 3 NRs. For the conduction states with $E>E_c$, LDA-1/2 and mBJ match \GW{} better than LDA. It is also apparent that, the peaks in the energy range 0-5 eV are more pronounced in LDA-1/2 than in other methods.

The definition of $E_v$ and $E_c$ allows us to compute $\Delta E$ as
\begin{equation}
	\Delta E = E_c - E_v.
\end{equation}
$\Delta E$ can be identified as a kind of band gap for the core region of the NR, since it is obtained from band edges of states that belong to core atoms.

Table \ref{tab:gaps-nanorods} presents $\Delta E$ for each NR.
The best agreement with \GW{} is given by LDA-1/2, with $\Delta E$ approximately 1.2 eV smaller. 
\begin{table}[htb]
	\caption{$\Delta E$ of core-shell In$_n$Al$_{38-n}$N$_{38}$H$_{40}$, obtained from band edges of DOS projected onto atoms in the NR core.}\label{tab:gaps-nanorods}
	\begin{tabular}{cccc}
		\hline
		& & $n$ \\
		& 0 & 2 & 4 \\ \hline
		\GW{} & 8.44 & 8.38 & 8.14\\
		LDA & 4.69 & 4.76 & 4.75 \\
		LDA-1/2 & 7.26 & 7.19 & 7.00 \\
		mBJ & 6.66  & 6.71 & 6.55 \\
		\hline
	\end{tabular}
\end{table}
mBJ comes next, predicting $\Delta E$ 1.6-1.8 eV smaller than \GW. LDA is the last one with $\Delta E$ 3.4-3.7 eV smaller than \GW.

When compared to bulk AlN, the NRs are expected to have larger $\Delta E$ due to quantum confinement effects. 
Indeed, the passivated AlN NR has $\Delta E$ larger than bulk AlN by 2.15, 0.45, 1.05 and 0.96 and eV, when calculated with \GW, LDA, LDA-1/2 and mBJ respectively. 
As estimated in Appendix \ref{appendix:bandgap}, an enlargement of 1.9-2.5 eV is expected due to quantum confinement effects. 
\GW{} best matches this expectation, followed by LDA-1/2, mBJ and LDA.

\subsubsection{Optical properties}
\begin{figure}[htb]
	\includegraphics[scale=1]{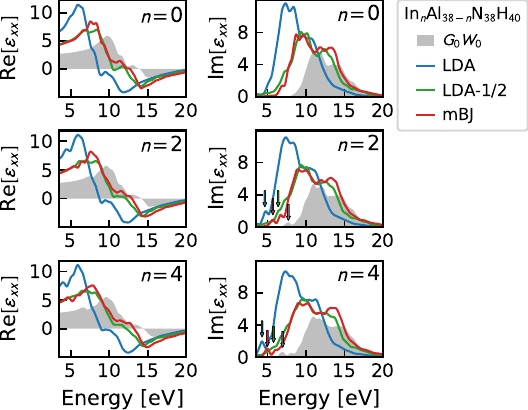}	
	\caption{$xx$ component of the dielectric function, with its real and imaginary parts. The label $n$ refers to the amount of In atoms in the NR cell according to In$_n$Al$_{38-n}$N$_{38}$H$_{40}$. For NRs with $n>0$, the arrows point peaks coming from In contributions.}\label{fig:epsilon-nanorods}
\end{figure}

Figure \ref{fig:epsilon-nanorods} displays the $xx$ component of the dielectric function. 
LDA-1/2 and mBJ agree well with each other and are red-shifted by 1.8 and 2.0 eV, respectively, in comparison with \GW{}. For LDA, this amounts to 3.5 eV. By blue-shifting all Im$[\varepsilon_{xx}]$, a good agreement with \GW{} can be observed, as shown in Appendix \ref{appendix:shift-epsilon}. Although these shifts do not reproduce exactly the differences in $\Delta E$, they are comparable.

Next, we consider the contribution of In atoms present in the core region to the dielectric function of the NRs. 
In Fig. \ref{fig:epsilon-nanorods}, it is evident that the presence of In introduces peaks in $\mathrm{Im}[\varepsilon_{xx}]$, which are red-shifted in respect to the main peak observed for AlN NRs without In.
These peaks, highlighted with arrows in Fig. \ref{fig:epsilon-nanorods}, become evident when we $\mathrm{Im}[\varepsilon_{xx}]$ for NRs with In and by $\mathrm{Im}[\varepsilon_{xx}]$ for AlN NRs (not shown here). Table \ref{tab:peaks-nanorods} shows the positions of these peaks due to In.
\begin{table}[htb]
	\caption{Peak positions due to In atoms in core-shell In$_n$Al$_{38-n}$N$_{38}$H$_{40}$ NRs. These peaks are highlighted in Fig. \ref{fig:epsilon-nanorods} with arrows. }\label{tab:peaks-nanorods}
	\begin{tabular}{ccc}
		\hline
		& \multicolumn{2}{c}{$n$} \\
		&  2 & 4 \\ \hline
		\GW{} & 7.70 & 6.96  \\
		LDA & 4.64 & 4.28 \\
		LDA-1/2 & 6.37 & 5.77 \\
		mBJ & 5.69  & 4.97 \\
		\hline
	\end{tabular}
\end{table}
Peaks within \GW{} appear blue-shifted in comparison with other methods.
Peaks obtained with LDA-1/2 exhibit best agreement with \GW, with a difference of 1.2-1.3 eV. These numbers are 2.0 and 3.1-3.4 eV for mBJ and LDA, respectively. Also interesting is that the red-shift of the peaks observed by increasing $n=2$ to $n=4$ is approximately the same in \GW{} (0.74 eV), LDA-1/2 (0.60 eV) and mBJ (0.72 eV).

\begin{figure}[htb]
	\includegraphics[scale=1]{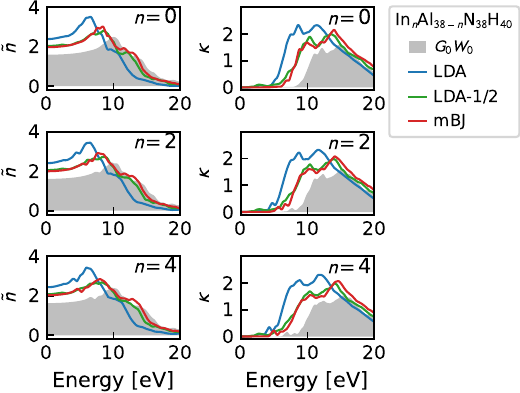}	
	\caption{Refraction index $\tilde{n}$ and extinction coefficient $\kappa$.}\label{fig:refraction-index}
\end{figure}

In Fig. \ref{fig:refraction-index}, we depict the refraction index $\tilde{n}$ and the extinction coefficient $\kappa$ for the energy range of 0 to 20 eV. 
The similarity between LDA-1/2 and mBJ is apparent. Regarding the refraction index, for the energy range 10-20 eV, LDA-1/2 and mBJ show an excellent agreement with \GW. Although this statement does not hold for the extinction coefficient, there is a notable improvement over LDA: LDA-1/2 and mBJ approximate \GW{} better than LDA.

Table \ref{tab:refractive-index} presents the static refractive index. The best agreement with respect to \GW{} is observed for mBJ (difference of 12-13\%), closely followed by LDA-1/2 (14\%) and, then, by LDA (26-27\%). 

\begin{table}[htb]
	\caption{Refractive index $\tilde{n}$ at zero frequency of core-shell In$_n$Al$_{38-n}$N$_{38}$H$_{40}$ NRs. }\label{tab:refractive-index}
	\begin{tabular}{cccc}
		\hline
		& \multicolumn{3}{c}{$n$} \\
		&  0 & 2 & 4 \\ \hline
		\GW{} & 1.59 & 1.61 & 1.63  \\
		LDA & 2.38 & 2.40 & 2.43\\
		LDA-1/2 & 2.03 & 2.05 & 2.07 \\
		mBJ & 1.97  & 1.99 & 2.02 \\
		\hline
	\end{tabular}
\end{table}

\begin{figure}[htb]
	\includegraphics[scale=1]{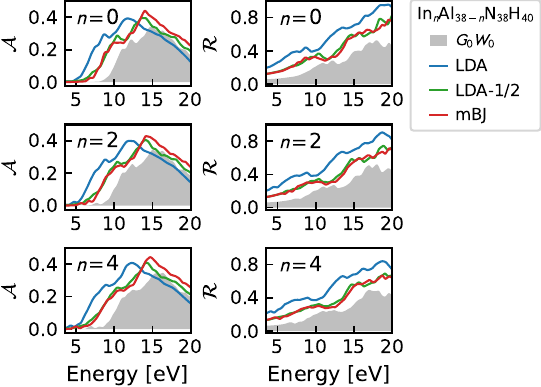}	
	\caption{Absorption $\mathcal{A}$ and reflectance $\mathcal{R}$.}\label{fig:absorption-reflectance}
\end{figure}

Figure \ref{fig:absorption-reflectance} depicts the absorbance and the reflectance of the NRs. The curves for LDA-1/2 and mBJ are very similar, and both present a better agreement with \GW{} than LDA.

%\begin{figure}[htb]
%	\includegraphics[scale=1]{figs/deltaepsilon-nanorods.pdf}	
%	\caption{Imaginary part of dielectric function.}\label{fig:deltaepsilon}
%\end{figure}

\section{Conclusion}\label{sec:conclusions}
We have studied electronic and optical properties of core-shell InAlN NRs with LDA, LDA-1/2, mBJ and \GW. For the properties, DOS, dielectric function, refractive index, extinction coefficient, absorption coefficient, and reflectance: results with LDA-1/2 and mBJ are similar and agree better with \GW{} than those obtained with LDA. For band gaps and peaks in Im$[\varepsilon]$ coming from In contributions, LDA-1/2 agrees better with \GW{} than mBJ. Overall, LDA-1/2 and mBJ can be used as tools to replace \GW{} with reasonable accuracy at much less computational cost. 
\par The authors have no conflicts to disclose. The data that support the findings of this study are available from the corresponding author upon reasonable request.

\acknowledgements{}
The authors gratefully acknowledge the computing time granted by the Resource Allocation Board and provided on the supercomputer Lise and Emmy at NHR@ZIB and NHR@Göttingen as part of the NHR infrastructure. 
They also acknowledge resources provided by the National Academic Infrastructure for Supercomputing in Sweden (NAISS) at the National Supercomputer Center (NSC) in Linköping (NAISS 2023/5-116 and NAISS 2023/23-161) partially funded by the Swedish Research Council through grant agreement no. 2018-05973.
G. K. G., J. B., and L. H. acknowledge support by the Swedish
Government Strategic Research Area in Materials Science on Advanced Functional Materials (AFM) at Linköping University (Faculty Grant SFO-Mat-LiU No. 2009-00971). 
C.-L.H. acknowledges support by the Swedish Research Council (Vetenskapsrådet) through grant number 2018-04198 and the Swedish Energy Agency (Energimyndigheten) through grant
number 46658-1. 

\bibliography{references}

\section*{Appendixes}
\appendix
\section{Convergence behavior}\label{appendix:convergence}

\subsection{Bulk AlN and InN}\label{appendix:convergence-bulk}
\begin{figure}[htb]
	\includegraphics[scale=1]{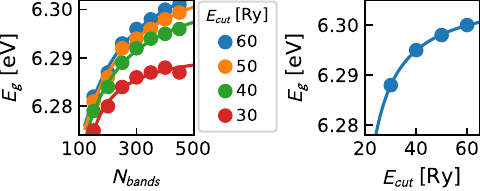}	
	\caption{AlN band gap obtained with \GW: convergence behavior with respect to the number of bands (left) and the planewave cutoff for the dielectric function (right).}\label{fig:AlN-GW-conv}
\end{figure}

\begin{figure}[htb]
	\includegraphics[scale=1]{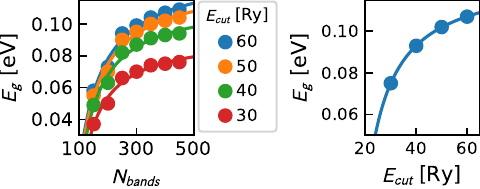}	
	\caption{Same as Fig. \ref{fig:AlN-GW-conv} for InN.}\label{fig:InN-GW-conv}
\end{figure}

Figures \ref{fig:AlN-GW-conv} and \ref{fig:InN-GW-conv} show, for bulk AlN and InN, the convergence behavior of the \GW{} band gap. Two parameters are varied: the number of KS states ($N_{bands}$) and the planewave cutoff ($E_{cut}$) used to build the dielectric function.
The convergence of \GW{} band gap is slower than that of DFT. Having a fully converged band gap is challenging, as employing a set of fully converged parameters comes with a very high computational cost. Therefore, we follow here an extrapolation procedure to evaluate the \GW{} band gap, adopting the expression\cite{Klimes_2014,Nabok_2016,Pela_2016}:
\begin{equation}
	E_g(x) = E_g(\infty) + \frac{A}{x+B},
\end{equation}
where $E_g(\infty)$, $A$ and $B$ are fit coefficients, and $x$ is the convergence parameter being tested, which can be $N_{bands}$ or $E_{cut}$. Assuming that the extrapolation with respect to $E_{cut}$, $N_{bands}$ and the k-grid can be carried out separately, the extrapolated band gap $E_g^{extr}$ can be evaluated as
\begin{equation}
	E_g^{extr} = E_g^{ref} + \delta_{kpt} + \delta_{bands} + \delta_{cut},
\end{equation}
where $E_g^{ref}=E_g(N_{bands}^{ref},E_{cut}^{ref},k_{grid}^{ref})$ is the \GW{} band gap for a reference calculation employing $N_{bands}^{ref}$, $E_{cut}^{ref}$, and $k_{grid}^{ref}$. We adopt for AlN: $N_{bands}^{ref}=200$, $E_{cut}^{ref}=40$ Ry; and for InN: $N_{bands}^{ref}=400$, $E_{cut}^{ref}=50$ Ry; and, in both cases, $k_{grid}^{ref}$ as $4\times 4 \times 3$.

Then $\delta_{cut}$ and $\delta_{bands}$ are obtained as
\begin{equation}
	\delta_{bands} = E_g(N_{bands}=\infty,E_{cut}^{ref},k_{grid}^{ref})-E_g^{ref},
\end{equation}
\begin{equation}
	\delta_{cut} = E_g(N_{bands}^{ref},E_{cut}=\infty,k_{grid}^{ref})-E_g^{ref},
\end{equation}
and the contribution of the k-grid to the extrapolation is approximated as\cite{Klimes_2014,Nabok_2016,Pela_2016}:
\begin{equation}
	\delta_{kpt} = E_g(N_{bands}^{ref},E_{cut}^{ref},k_{grid}^{large})-
	E_g(N_{bands}^{ref},E_{cut}^{ref},k_{grid}^{ref}),
\end{equation}
where $8\times 8 \times 6$ is employed as $k_{grid}^{large}$.

Table \ref{tab:extrapolated_bandgaps} collects the relevant data regarding the extrapolation of the \GW{} band gap.
\begin{table}[htb]
	\centering
	\caption{Contributions to the extrapolated \GW{} band gap. All quantities are given in eV.}\label{tab:extrapolated_bandgaps}
	\begin{tabular}{cccccc}
	\hline
	 & $E_g^{ref}$ & $\delta_{bands}$ & $\delta_{cut}$ & $\delta_{kpt}$ & $E_g^{extr}$ \\ \hline
	 AlN & 6.28 & 0.03 & 0.01 & -0.03 & 6.29\\
	 InN & 0.09 & 0.02 & 0.03 & 0.14 & 0.28\\
	 \hline
	\end{tabular}
\end{table}

\subsection{Nanorods}
\subsubsection{Vacuum}
In Eq. (\ref{eq:im_epsilon}), $\Omega$ stands for the volume of the sample. 
However, in most of \emph{ab initio} codes, including Quantum Espresso, $\Omega$ is treated as the unit cell volume. For systems with vacuum in the unit cell, such as the NRs studied here, the dielectric function must be corrected by a factor $h$ equal to the ratio between the volumes of the cell and the sample. In terms of the geometry shown in Fig. \ref{fig:Shell-Core}, we have for the NRs
\begin{equation}
	h = \frac{A_{cell}}{A_{nanorod}} = \frac{4L^2}{3d^2},
\end{equation}
where $A_{cell}$ and $A_{nanorod}$ are the cross-sections of the unit cell and the NR, respectively.

Denoting $\varepsilon'_{xx}$ as the dielectric function without the correction and $\varepsilon_{xx}$, the corrected one, then according to Eqs. (\ref{eq:im_epsilon}) and (\ref{eq_re_epsilon}), it follows that:
\begin{equation}
	\mathrm{Im[\varepsilon_{xx}]} = h\mathrm{Im[\varepsilon'_{xx}]},
\end{equation}
\begin{equation}
	\mathrm{Re[\varepsilon_{xx}]} = 1 + h(\mathrm{Re[\varepsilon'_{xx}]}-1).
\end{equation}

\begin{figure}[htb]
	\includegraphics[scale=1]{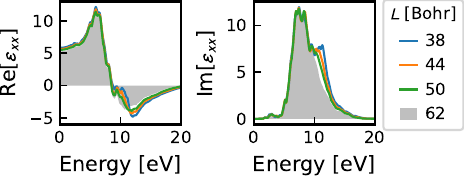}	
	\caption{Convergence behavior of the AlN NR - dielectric function obtained with LDA.}\label{fig:AlN-conv-vacuum}
\end{figure}

Figure \ref{fig:AlN-conv-vacuum} depicts the convergence behavior of the dielectric function with respect to cell dimension $L$.
These calculations refer to the AlN NR and have been carried out with LDA, and a similar trend has been observed for the other cases. Fig. \ref{fig:AlN-conv-vacuum} shows that the size $L=44$ Bohr, adopted for the NRs calculations reported in the present work, is sufficient to guarantee satisfactory convergence within the energy range of $0-20$ eV.

\subsubsection{Band gap}
Here, we discuss the precision level expected for our \GW{} calculations concerning the NRs. As for bulk AlN and InN, we check convergence with respect to $N_{bands}$, $E_{cut}$ and the k-grid $1\times 1 \times N_{kpt}$. For the sake of computational cost, we restrict the analysis to the $\Gamma\Gamma$ band gap and the dielectric function of AlN.

\begin{figure}[htb]
	\includegraphics[scale=0.95]{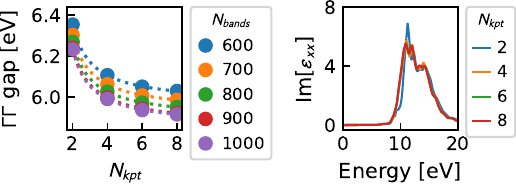}	
	\caption{Convergence behavior of \GW{} results for the AlN NR: $\Gamma\Gamma$ band gap (left) and dielectric function (right).}\label{fig:AlN-conv-kpt}
\end{figure}

Figure \ref{fig:AlN-conv-kpt} depicts on the left side the impact of $N_{kpt}$ and $N_{bands}$ on the $\Gamma\Gamma$ band gap. Since we adopted $N_{kpt}=6$ and $N_{bands}=900$, we estimate, by making the extrapolation procedure as described in Appendix \ref{appendix:convergence-bulk}, a correction of $\delta_{bands}=-0.06$ eV and $\delta_{kpt}=-0.08$ eV.

On the right side of Fig. \ref{fig:AlN-conv-kpt}, we present the imaginary part of dielectric function. Even though the band gap has a more pronounced dependence on $N_{kpt}$, the dielectric function exhibits good convergence already for $N_{kpt}=6$.

\begin{figure}[htb]
	\includegraphics[scale=1]{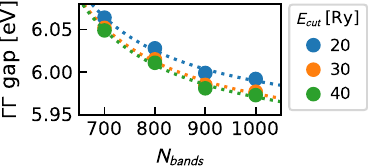}	
	\caption{Impact of the number of bands on $\Gamma\Gamma$ band gap of AlN NR.}\label{fig:AlN-conv-bands}
\end{figure}

Figure \ref{fig:AlN-conv-bands} displays the influence of $N_{bands}$ and $E_{cut}$ on the $\Gamma\Gamma$ band gap. Following the extrapolation procedure, we evaluate a correction of $\delta_{cut}=-0.03$ eV for the adopted $E_{cut}=20$~Ry. Overall, by adding up $\delta_{bands}$, $\delta_{kpt}$ and $\delta_{cut}$, we expect that \GW{} band gaps are overestimated by about 0.2 eV.

\section{Quantum confinement and band gap}\label{appendix:bandgap}
The expected band gap enlargement $\Delta$ due to quantum confinement effects can be estimated by considering a particle of mass $\mu$ confined inside a circular quantum well of radius $r$. Following Ref. \onlinecite{Robinett_1996}, we have, in atomic units:
\begin{equation}\label{eq:DeltaBandGap}
\Delta = \frac{a^2_{(0,0)}}{2\mu^*r^2},
\end{equation}
where $a_{(0,0)}=2.40483$ is the first zero of $J_0(x)$, Bessel's function of order zero.
To estimate $\Delta$ for the case of the NRs, we approximate $\mu$ by the reduced mass of the electron and the heavy hole, $\mu=(1/m_e + 1/m_{hh})^{-1}$. Taking the suggested values in Ref. \onlinecite{Vurgaftman_2003}: $m_e=\sqrt[3]{(m_e^\perp)^2 m_e^\parallel}=0.31$, in atomic units. $m_{hh}$ is obtained from the Pikus-Bir parameters given in Ref. \onlinecite{Vurgaftman_2003} as\cite{Chuang_1996}
\begin{equation}
m_{hh}=-\sqrt[3]{(A_2+A_4)^{-2}(A_1+A_3)^{-1}},
\end{equation} 
which gives $m_{hh}=1.1$, in atomic units. Therefore $\mu=0.24$. 

Lastly, we consider a range for $r$ between the radius of the inscribed and the circumscribed circles in the NRs. Taking into account the NR diameter of $14$~\AA, this implies that $r$ lies between $6.06$ and $7$ \AA. Applying Eq. (\ref{eq:DeltaBandGap}) gives the range of $1.9$-$2.5$ eV for $\Delta$.

\section{Shift in the dielectric function}\label{appendix:shift-epsilon}
\begin{figure}[htb]
	\includegraphics[scale=1]{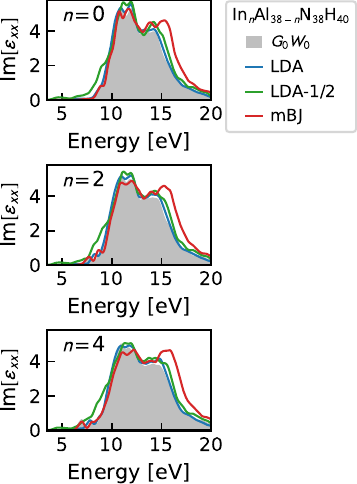}	
	\caption{In$_n$Al$_{38-n}$N$_{38}$H$_{40}$ core-shell NRs: Imaginary part of the dielectric function blue-shifted by $\delta$. LDA, LDA-1/2 and mBJ have been shifted by 3.5, 1.8 and 2.0 eV, respectively.}\label{fig:epsilon-nanorods-shift}
\end{figure}

Figure \ref{fig:epsilon-nanorods-shift} shows, for the NRs, $\mathrm{Im}[\varepsilon_{xx}]$ of the different DFT approaches blue-shifted by $\delta$ to better match \GW. To plot these curves, we shift the vertical transitions in Eq. (\ref{eq:im_epsilon}), so that the shifted dielectric function $\varepsilon'_{xx}$ at a given frequency $\omega$ relates to the original $\varepsilon_{xx}$ by:
\begin{equation}
\mathrm{Im}[\varepsilon'_{xx}(\omega)] = 
\mathrm{Im}[\varepsilon_{xx}(\omega-\delta)]\frac{(\omega-\delta)^2}{\omega^2}.
\end{equation}
It is observed that, setting $\delta$ to 3.5, 1.8 and 2.0 eV for LDA, LDA-1/2 and mBJ, respectively, leads to an excellent agreement with $\mathrm{Im}[\varepsilon_{xx}]$ obtained with \GW.
\end{document}